\documentclass{article}

\usepackage[a4paper, total={6in, 8in}]{geometry}

\usepackage{amsmath}

\usepackage{amssymb}

\usepackage{cite}

\title{Type V singularities in non-standard cosmological backgrounds}

\author{Oem Trivedi$^a$ \footnote{oem.t@ahduni.edu.in}  \and Maxim Khlopov \footnote{khlopov@apc.in2p3.fr}  $^{b,c,d}$}
\date{%
	$^a$School of Arts and Sciences, Ahmedabad University,Ahmedabad 380009,India\\%
	$^b$Institute of Physics, Southern Federal University,
	Stachki 194 Rostov on Don 344090, Russia\\%
	$^c$Universite de Paris, CNRS, Astroparticule et Cosmologie
	F-75013 Paris, France\\%
	$^d$National Research Nuclear University ”MEPHI” 115409 Moscow, Russia\\%
	\today
}

\begin{document}
	
	\maketitle
	
	\begin{abstract}
		Interest in cosmological singularities has remarkably grown in recent times, particularly on future singularities with the discovery of late-time acceleration of the universe and dark energy. While there has been an expansive literature which has discussed the occurrence of Type I-IV singularities in many non-standard cosmologies, w/Type V singularities have not yet been explored in exotic cosmological settings. So in this work we pursue the same and discuss the occurrence of w-singularities in a variety of non-standard cosmologies. We consider the RS-II Braneworld cosmology, an F(R) gravity cosmology which gives viable late time acceleration and we also consider cosmologies due to modified area-entropy relations, generalized uncertainty principles, holographic renormalization and Chern-Simons gravity( all of which can be coincidentally described by the same form of the modified Friedmann equation). We show that w-singularities will occur in exactly the same conditions in all these cosmological settings as they do in the usual general relativistic cosmology if one considers a power series expansion ansatz for the scale factor. We also show that if one considers an exponential form of the scale factor then while Type V singularities in the RS-II Braneworld and Chern-Simons cosmologies occur in the same conditions as in the standard general relativistic case, there is a significant difference in the conditions when one considers the f(R) gravity case. These results are surprising overall, as one would usually not expect cosmological singularities to occur in almost the same conditions in non-standard cosmologies as they do in the usual standard cosmology.
	\end{abstract}


	\maketitle
	
	\section{Introduction }
	
	Observations of late time acceleration of the Universe surprised the whole cosmological community \cite{SupernovaSearchTeam:1998fmf} . Ever since then a lot of work has been done in order to explain this expansion which include the standard ones like the Cosmological constant \cite{SupernovaSearchTeam:1998fmf,Weinberg:1988cp,Lombriser:2019jia,Copeland:2006wr,Padmanabhan:2002ji} alongside more exotic scenarios like Modified gravity theories\cite{Capozziello:2011et,Nojiri:2010wj,Nojiri:2017ncd}and very appealing ways of detecting dark energy directly have been put forward recently as well \cite{Zhang:2021ygh}. An exciting approach towards understanding dark energy is that of Quintessence, where a scalar field drives the late-time cosmic acceleration of the universe\cite{Zlatev:1998tr,Tsujikawa:2013fta,Faraoni:2000wk,Gasperini:2001pc,Capozziello:2003tk,Capozziello:2002rd,Carroll:1998zi,Caldwell:2005tm,Han:2018yrk,Astashenok:2012kb,Shahalam:2015sja}.There has also been an expansive literature in recent times which has been devoted to the study of various types of singularities that could occur during the current and far future of the Universe with emphasis to various approaches towards dark energy \cite{Nojiri:2004ip,Nojiri:2005sr,Nojiri:2005sx,Bamba:2008ut}. These cosmological singularities which have been discussed in recent times can be classified broadly into two types ; strong and weak. 
	\\
	\\
	Strong singularities are those singularities which can distort finite objects and can mark either the beginning or the end of the universe, with the big bang being the one for the start of the universe and the so called "big rip" signalling the end of the universe. Weak singularities, as the name might suggest, are those which do not have such far reaching implications and do not distort finite objects in the same sense as their strong counterparts. We can discuss these various singularities in more detail as follows : \begin{itemize}
		\item Type I ("Big Rip") : In this case, the scale factor , effective energy density and effective pressure density diverges. This is a scenario of universal death, wherein everything which resides in the universe is progressively torn apart \cite{Caldwell:2003vq}.
		\item Type II ("Sudden/Quiescent singularity") : In this case, the pressure density diverges and so does the derivatives of the scalar factor from the second derivative onwards \cite{Barrow:2004xh}. The weak and strong energy conditions hold for this singularity. Also known as quiescent singularities, but this name originally appeared in contexts related to non-oscillatory singularities \cite{Andersson:2000cv}. A special case of this is the big brake singularity \cite{Gorini:2003wa}.
		\item Type III ("Big Freeze") : In this case, the derivative of the scale factor from the first derivative onwards diverges. These were detected in generalized Chaplygin gas models \cite{bouhmadi2008worse}.
		\item Type IV ("Generalized sudden singularities"): These are finite time singularities with finite density and pressure instead of diverging pressure. In this case, the derivative of the scale factor diverges from a derivative higher than the second \cite{Bamba:2008ut}.
		\item Type V ("w-singularities") : In this case, the scale factor, the energy and pressure densities are all finite but the barotropic index $w = \frac{p}{\rho}$ becomes singular, first shown in \cite{Dabrowski:2009kg} (also investigated interestingly in \cite{Fernandez-Jambrina:2010ngm}) . 
		\item Type $\infty$(" Directional singularities "): Curvature scalars vanish at the singularity but there are causal geodesics along which the curvature components diverge \cite{Fernandez-Jambrina:2007ohv} and in this sense, the singularity is encountered just for some observers. 
		\item Inaccessible singularities: These singularities appear in cosmological models with toral spatial sections, due to infinite winding of trajectories around the
		tori. For instance, compactifying spatial sections of the de Sitter model to cubic tori. However, these singularities cannot be reached by physically well defined observers and hence this prompts the name inaccessible singularities \cite{mcinnes2007inaccessible}.
	\end{itemize}
	The study of dark energy in various non GR based backgrounds have gathered a deep interest in recent times, particularly as several such theories have distinct features which are not present in the usual dark energy regimes based in GR. These cosmological backgrounds are very diversely motivated themselves and have their origins in various. For example,a lot of works have considered the possibility of viable scalar field based dark energy regimes in quantum gravity corrected cosmologies like the RS-II Braneworld and Loop Quantum Cosmology \cite{Sahni:2002dx,Sami:2004xk,Tretyakov:2005en,Chen:2008ca,Fu:2008gh}. There has been substantial work on new dark energy models based on thermodynamic modifications like modified area-entropy relations as well\cite{Tavayef:2018xwx,Radicella:2011qpl,Bamba:2009id,Younas:2018kmy,Jawad:2016tne,Nojiri:2019skr}. A great amount of literature has also been devoted to look out for exotic dark energy regimes based in cosmologies where the generalized uncertainty principles \cite{maggiore1994quantum,Adler:2001vs,tawfik2014generalized,Barca:2021epy} are considered instead of the usual Heisenberg uncertainty criterion \cite{Ghosh:2011ft,Rashki:2019mde,Paliathanasis:2021egx}. There has been a lot of work devoted towards studying dark energy regimes in non-canonical approaches like DBI etc. as well \cite{Calcagni:2006ge,Gumjudpai:2009uy,Chiba:2009nh,Ahn:2009xd,Li:2016grl,mandal2021dynamical,Kar:2021gbz}. 
	\\
	\\
	This vast dark energy literature has prompted the study of cosmological singularities in a wide range of cosmological backgrounds as well, as there have been multiple works which have discussed Type I-IV singularities in various cosmologies \cite{Shtanov:2002ek,Bamba:2012ka,Bamba:2010wfw,Nojiri:2008fk,odintsov2018dynamical,Odintsov:2018awm,trivedi2022finite,Bombacigno:2021bpk,Nojiri:2006gh,Fernandez-Jambrina:2021foi,Chimento:2015gga,Chimento:2015gum,Cataldo:2017nck}. But Type V or w singularities have not been studied in the context of cosmological backgrounds which are not primarily general relativistic. At this point, a discussion is in order over the cosmological significance of w-singularities. While Type I-Type IV singularities deal with more direct cosmological parameters like the scale factor, Hubble parameter alongside energy and pressure densities, type V singularities deal with a somewhat indirect parameter in the form of w. This is not say, however, that these singularities cannot occur in cosmological and in particular, dark energy models. For example \cite{Elizalde:2018ahd}  discussed how w-singularities can occur in interacting dark energy models(while the background cosmology in this case was still general relativistic and the continuity equation had its usual form), while in \cite{Khurshudyan:2018kfk} it was showed how varying Chaplygin gas models can also have w-singularities. The occurrence of w-singularities in various other contexts has also been discussed in \cite{Szydlowski:2017evb,Samanta:2017qnn,Sadri:2018lzz,Astashenok:2012tv,ozulker2022dark}. Hence while type V singularities deal primarily with a more indirect cosmological parameter, it by no means diminishes its cosmological importance and it does appear in a variety of cosmological expansion scenarios. This encourages one towards an endeavour which tries to understand the status quo of w-singularities in non-standard cosmologies, which we will be undertaking here. We will consider cosmologies which have a wide range of motivations, ranging from String theory to modified gravity theories, and we will explore under which conditions w-singularities can occur ( if they even will) in these scenarios and if these conditions are any different than what happens in the standard general relativistic case. In Section II, we will be briefly describing the cosmologies we want to consider here and in section III we will work out the conditions for w-singularities to occur in these scenarios. We will finally conclude our work in section IV.
	\section{A brief review of the cosmologies considered}
	In order to be the as general as we can in our approach, we should try to take into account cosmologies which have been modified from the usual GR behaviour due to a variety of effects. To this end we will hence be considering cosmologies which get their distinguishing features from GR due to quantum gravity corrections, F(R) and Chern Simons gravity modifications, modified area-entropy relations, holographic renormalization and generalized uncertainty principles. We will briefly describe about the primary changes in the cosmological dynamics which will be crucial to our work here, starting with the RS-II Braneworld. 
	\\
	\\
	The RS-II model is a based on a modification of the RS-I Braneworld cosmology model\cite{Randall:1999ee}, where the hierarchy problem is solved by embedding two 3-branes in a five-dimensional bulk where one of the branes contains the Standard Model particles. The RS-II braneworld cosmology removes one of the 3-branes and recovers both Newtonian gravity and General Relativity as its limiting cases\cite{Randall:1999vf}. Since another braneworld cosmology scenario in the form of Dvali-Gabadadze-Porrati (DGP)  model \cite{Deffayet:2000uy,Dvali:2000hr} can produce some effects on the late-time evolution of the universe, it becomes very interesting to see if one can address late-time acceleration issues using the RS-II model as well. In this direction, there have been several works in recent times which have addressed  Quintessence in the RS-II Braneworld scenario\cite{Sahni:2002dx,Sami:2004xk,Bento:2008yx}. We can write the total action of the RS-II Model inclusive of both the scalar and the background fluid term as \begin{equation}
	S  = S_{RS} + S_{B} + S_{\phi} = \int d^5 x \sqrt{-g^{(5)}} \left( \Lambda^{(5)}  + 2 R^{(5)}   \right) + \int d^4 x \sqrt{-g} \left(\lambda -\frac{1}{2} \mu(\phi) (\nabla \phi)^2 - V(\phi)  + \mathcal{L}_{B}  \right) 
	\end{equation}  where $R^{(5)} $, $ g^{(5)}_{\mu \nu} $ and $ \Lambda^{(5)} $ are the bulk Ricci Scalar, metric and the cosmological constant respectively with $\lambda$ being the brane tension on the 3-brane, $g_{\mu \nu} $ being the 3-brane metric and $\mu(\phi) $ being a scalar coupling function. Note that here we are working in Planck units with $(m_{p}^{(5)})^2 = 1 $ with $ m_{p}^{(5)} $ being the 5-dimensional Planck mass. Assuming that the brane metric has the usual FLRW form, we get the Friedmann equation to be \cite{Maartens:2010ar} \begin{equation}
	H^{2}  = \rho \left(1 + \frac{\rho}{2 \lambda}\right)
	\end{equation} 
	where $ \rho = \rho_{\phi} + \rho_{B} $ is the total cosmological energy density taking into account contributions from both the scalar field and the background fluid term and the Bulk cosmological constant has been set to zero for simplicity. The equation motion of the scalar is given by \begin{equation}
	\mu(\phi) \ddot{\phi} + \frac{1}{2} \frac{d \mu}{d \phi} \dot{\phi}^2  + 3 H \mu(\phi) \dot{\phi} + \frac{dV}{d\phi} = 0 
	\end{equation}
	As we are interested in simplified cases for the RS-II Braneworld, we will consider the coupling $ \mu = 1 $ but besides this we do not place any constraints on the brane tension $\lambda$ or on the potential V$(\phi)$ hence our analysis still is very general.
	\\
	\\
	The $F(R)$ gravity scenario that we would like to consider has the action \cite{Carroll2004} \begin{equation}
	S = \frac{m_{p}^2}{2} \int d^4 x \sqrt{-g} \left(R - \frac{\alpha^2}{R}\right) + \int d^4 x \sqrt{-g} \mathcal{L}_{m}
	\end{equation}
	where $\alpha$ is a constant which has the units of mass, $\mathcal{L}_{m} $ is the Lagrangian density for matter and $m_{p}$ is the reduced planck's constant. The field equation for this action is \begin{equation}
	\left(1 + \frac{\alpha^2}{R^2}\right) R_{\mu \nu} - \frac{1}{2} \left(1 - \frac{\alpha^2}{R^2}\right) R g_{\mu \nu} +  \alpha^2 \left[g_{\mu \nu} \nabla_{a} \nabla^{a} - \nabla (_{\mu} \nabla_{\nu})  \right] R^{-2} = \frac{T_{\mu \nu}^{M}}{m_{p}^2}
	\end{equation}
	where $T_{\mu \nu}^{M}$ is the matter energy-momentum tensor. The Friedmann equation in this case can take the form \begin{equation}
	\frac{6 H^2 - \frac{\alpha}{2}}{11/8 - \frac{8 H^2}{4 \alpha}} = \frac{\rho}{3}
	\end{equation} where $\rho$ is the total energy density. This $F(R)$ gravity regime was used to explain late time cosmic acceleration as an alternative to dark energy in \cite{Carroll2004}. The action prompts one towards the notion that very tiny corrections to the usual Einstein Hilbert in the form of $R^{n}$ with $n<0$ can produce cosmic acceleration. As corrections of the form $R^n$ with $n>0$ can lead to inflation in the early universe \cite{Starobinsky:1980te}, the authors in \cite{Carroll2004} proposed a purely garvitational paradigm through (5) to explain both the early and late time accelerations of the universe. Explaining the current epoch of the universe through such a modified gravity model would in principle eliminate the need of dark energy and hence, its an interesting scenario to consider w-singularities in as well.
	\\
	\\
	The third modified Friedmann equation that we will be covering is given by \begin{equation}
	H^{2} - \alpha H^{4} = \frac{\rho}{3}
	\end{equation}
	The above equation is very special in the sense that it can be derived from various distinct approaches. This equation can be achieved by considering a quantum corrected entropy-area relation of $ S = \frac{A}{4} - \alpha \ln \left( \frac{A}{4} \right) $ where A is the area of the apparent horizon and $\alpha$ is a dimensionless positive constant determined by the conformal anomaly of the fields, where the conformal anomaly is interpreted as a correction to the entropy of the apparent horizon \cite{cai2008corrected} . Also, this could be derived in terms of spacetime thermodynamics together with a generalized uncertainly principle of quantum gravity\cite{lidsey2013holographic}. This Friedmann equation can also be derived by considering an anti-de Sitter-Schwarzschild black hole via holographic renormalization with appropriate boundary conditions \cite{apostolopoulos2009cosmology}. Finally, a Chern-Simons type of theory can also yield this Friedmann equation \cite{gomez2011standard}. Hence the equation (7) can derived by a wide range of approaches to gravitational physics and can be representative of the effects of these different theories on the cosmological dynamics. In this way, besides the RS-II Braneworld and F(R) gravity theory discussed before, we are taking into account effects on cosmology from modified area-entropy relations, generalized uncertainty principles and holographic renormalization as well besides Chern-Simons gravity as well.
	\section{Condition for the occurrence of w-singularities}
	Barotropic index singularities or simply w-singularities were proposed by Fernandez-Jambrina in \cite{Fernandez-Jambrina:2010ngm}, where they showed that the barotropic index $ w = \frac{p}{\rho} $ can become singular if the scale factor can be represented by the expansion \begin{equation}
	a(t) = c_{0} + c_{1} (t_{s} - t)^{n1} + c_{2} (t_{s} - t)^{n2}......
	\end{equation} 
	where $t_{s} $ is the time of the singularity. In order for pressure to be finite, $ n_{1} > 1 $. Note that w-singularities were shown to form only when the scale factor takes up such power series forms, because if one considers a simple exponential scale factor like $a(t) = e^{\frac{b}{(t_{s} - t)^p}} $ with $p>0$ then w takes the form \begin{equation}
	w \approx -1 - \frac{2(p+1)}{3bp} (t_{s} - t)^p 
	\end{equation}
	which can never be singular \footnote{It can,however, produce other types of singularities and also there is a different form of the exponential scale factor which can give Type V singularities, which we will discuss later}. What we would like to do here is to check the whether the expansion (8) when considered in the generalized cosmologies discussed in Section 2 would still produce w-singularities and if so, will it form in the same conditions as in the general relativistic case. In order to do so, we would have to firstly express the energy density and pressure density completely in terms of the scale factor. The key to this endeavour is the continuity equation \begin{equation}
	\dot{\rho} + 3H(\rho + p) = 0
	\end{equation}
	which remains the same irrespective of what is the background cosmology. Considering the RS-II Friedmann equation (2), we can write the energy density in terms of the scale factor $a(t) $ as \begin{equation}
	\rho = \frac{\sqrt{\lambda a(t)^2 \left(6 a'(t)^2+\lambda a(t)^2\right)}}{a(t)^2}-\lambda
	\end{equation}
	where the overprimes denote derivatives with respect to time. Using the continuity equation (10), we can write the pressure density in this case as  \begin{equation}
	p = \frac{\lambda \left(-2 a(t) a''(t)+\sqrt{\lambda a(t)^2 \left(6 a'(t)^2+\lambda a(t)^2\right)}-4 a'(t)^2-\lambda
		a(t)^2\right)}{\sqrt{\lambda a(t)^2 \left(6 a'(t)^2+\lambda a(t)^2\right)}}
	\end{equation}
	And now we can write the barotropic index w as \begin{equation}
	w_{brane} = \frac{2 a(t) a''(t)-\sqrt{\lambda a(t)^2 \left(6 a'(t)^2+\lambda a(t)^2\right)}+4 a'(t)^2+\lambda a(t)^2}{\sqrt{\lambda a(t)^2 \left(6 a'(t)^2+\lambda a(t)^2\right)}-6 a'(t)^2-\lambda a(t)^2}
	\end{equation}
	Similarly we can write the energy density for the F(R) gravity scenario using the Friedmann equation (6) as \begin{equation}
	\rho = \frac{12 \alpha \left(\alpha a(t)^2-12 a'(t)^2\right)}{18 a'(t)^2 - 11 \alpha a(t)^2}
	\end{equation}
	And using the continuity equation, we can write the pressure density as \begin{equation}
	p = \frac{12 \left(216 \alpha a'(t)^4+11 \alpha^3 a(t)^4 -76 \alpha^2 a(t)^3 a''(t)\right)}{\left(18 a'(t)^2-11 \alpha a'(t)^2\right)^2}  - \frac{74 \alpha^2 a(t)^2 a'(t)^2 }{\left(18 a'(t)^2-11 \alpha a'(t)^2\right)^2}
	\end{equation}  After which we can write the w parameter in this case as \begin{equation}
	w_{f(R)} = \frac{\left(18 a'(t)^2-11 \alpha a(t)^2\right) \left(-76 \alpha a(t)^3 a''(t)-74 \alpha a(t)^2 a'(t)^2+\right)}{(18-11 \alpha)^2 a'(t)^4 \left(\alpha a(t)^2-12 a'(t)^2\right)} + \frac{216 a'(t)^4+11 \alpha^2 a(t)^4}{(18-11 \alpha)^2 a'(t)^4 \left(\alpha a(t)^2-12 a'(t)^2\right)}
	\end{equation}  Finally for the Friedmann equation (7), we can write the energy density as \begin{equation}
	\rho = \frac{3 a'(t)^2 \left(a(t)^2-n a'(t)^2\right)}{a(t)^4}
	\end{equation}
	After which we can again use the continuity equation and write the pressure density as \begin{equation}
	p = \frac{-4 \alpha a(t) a'(t)^2 a''(t) - 2 a(t)^3 a''(t)+ \alpha a'(t)^4+a(t)^2 a'(t)^2}{a(t)^4}  
	\end{equation}
	Finally we can write the barotropic index in this case \begin{equation}
	w_{CS} = \frac{2 a(t)^3 a''(t)+ \alpha a'(t)^4+a(t)^2 a'(t)^2-4 \alpha a(t) a'(t)^2 a''(t)}{3 \alpha a'(t)^4-3 a(t)^2
		a'(t)^2}
	\end{equation}
	where we have given the subscript "CS" to the w parameter to highlight that this is for the Chern-Simons cosmology( this Friedmann equation can be derived from a variety of approaches as discussed before but here we will refer to it just through Chern-Simons). Now we can use the ansatz (1) for the scale factor, from which one recovers very long expressions for each $w_{brane} $, $w_{f(R)}$ and $w_{CS}$ . Without reproducing those expressions, the central result that comes out is that one can only avoid w-singularities to occur in these cosmologies in the case that $ n_{1} \leq 1 $. This is the same result that one recovers in the case when one considers a general relativistic cosmology, as discussed in \cite{Fernandez-Jambrina:2010ngm}. This is a remarkable result because of the choice of cosmologies we have made here. The cosmological models we have studied through the Friedmann equations in (2,6,7) encapsulate not only f(R) gravity and String theoretic effects but also effects from modified area-entropy relations, holography, generalized uncertainty principle and Chern-Simons gravity. Hence if one did recover that the status quo of w-singularities is same in one of these regimes then perhaps it would not be very surprising, but seeing that even after considering loads of such effects and taking into account a lot of different cosmologies we still end up with the same conditions for avoiding w-singularities given the ansatz (1) is very fascinating.
	\\
	\\
	Another ansatz for the scale factor which can give w-singularities was proposed by Dabrowski and Marosek in \cite{Dabrowski:2012eb}, which has an exponential form in contrast to the power series form we have considered till now. That ansatz is given by \begin{equation}
	a(t) = a_{s} \left( \frac{t}{t_{s}}\right)^m \exp \left(1 - \frac{t}{t_{s}}\right)^{n}
	\end{equation}
	where $a_{s}$ has the units of length and is a constant while m and n are also constants \footnote{While the ansatz on the surface looks quite different from (8), it can be a sub case of (8) within certain limits as well } . The scale factor is zero (a=0) at t=0, thus signifying the big bang singularity. One can write the first and second derivatives of the scale factor as \begin{equation}
	\dot{a}(t) = a(t) \left[ \frac{m}{t} - \frac{n}{t_{s}} \left(1 - \frac{t}{t_{s}}\right)^{n-1}  \right] 
	\end{equation}
	\begin{equation}
	\ddot{a}(t)  = \dot{a}(t) \left[ \frac{m}{t} - \frac{n}{t_{s}} \left(1 - \frac{t}{t_{s}}\right)^{n-1}  \right] + a(t) \left[ - \frac{m}{t^2} + \frac{n(n-1)}{t^2} \left(1 - \frac{t}{t_{s}}\right)^{n-2}  \right] 
	\end{equation}
	where the overdots now denote differentiation with respect to time. From this, one can see that for $1<n<2$ $\dot{a} (0) \to \infty $ and $ \dot{a} (t_{s}) = \frac{m a_{s}}{t_{s}} = $ const. ,  while $a(t_{s}) = a_{s} $ , $\ddot{a}(0) \to \infty $ and $ \ddot{a} (t_{s}) \to -\infty $ and we have sudden future singularities. Furthermore, it was shown in \cite{Dabrowski:2012eb} that for the simplified case of the scale factor (20) with $m=0$, one can get w-singularities for $ n > 0 $ and $ n \neq 1 $. Note that this analysis was for a general relativistic and so we would now like to use the scale factor (20) for case $ m = 0 $, which can be written as \begin{equation}
	a(t) = a_{s} \exp \left(1 - \frac{t}{t_{s}}\right)^{n}
	\end{equation} and see if one gets w-singularities under similar conditions in the non-standard cosmological settings that we have considered here. Using the scale factor (23),  we can write the w-parameter for the Chern-Simons case (19) as
	\begin{equation}
	w_{CS} = -\frac{2 (n-1) n\alpha\left(1-\frac{t}{t_{s}}\right)^n}{3 n^2 \alpha\left(1-\frac{t}{t_{s}}\right)^{2 n}-3
		(t_{s}-t)^2}-\frac{2 (n-1) \left(1-\frac{t}{t_{s}}\right)^{-n}}{3 n}-1
	\end{equation} the barotropic index in this case is singular for all positive values of n besides $ n = 1 $, exactly like what happens in the usual general relativistic case. Hence even for the exponential ansatz (23), one does not see any difference in the conditions for the occurence of type V singularities from its standard cosmology case. This also remains true if one considers the RS-II Braneworld case (13), wherein w-singularities again occur for all positive values of n besides $n =1 $. For the case of f(R) gravity, however, things get more interesting. In that case, we can write the w parameter through (16) as   \begin{multline}
	w_{f(R)} = \frac{1368 (n-1) n^3 \alpha (t_{s}-t)^2 \left(1-\frac{t}{t_{s}}\right)^{3
			n} + 121 \alpha^3 (t_{s}-t)^6 }{n^4 (18-11 \alpha)^2 \left(12 n^2 \left(1-\frac{t}{t_{s}}\right)^{2 n}-\alpha
		(t_{s}-t)^2\right)} \\ + \frac{\left(1-\frac{t}{t_{s}}\right)^{-4 n} \left(-3888 n^6 \left(1-\frac{t}{t_{s}}\right)^{6 n}+5076 n^4 \alpha (t_{s}-t)^2 \left(1-\frac{t}{t_{s}}\right)^{4 n}\right)}{n^4 (18-11 \alpha)^2 \left(12 n^2 \left(1-\frac{t}{t_{s}}\right)^{2 n}-\alpha
		(t_{s}-t)^2\right)} \\ -  \frac{1848 n^2 \alpha^2 (t_{s}-t)^4 \left(1-\frac{t}{t_{s}}\right)^{2 n}+836 (n-1) n \alpha^2 (t_{s}-t)^4 \left(1-\frac{t}{t_{s}}\right)^n}{n^4 (18-11 \alpha)^2 \left(12 n^2 \left(1-\frac{t}{t_{s}}\right)^{2 n}-\alpha
		(t_{s}-t)^2\right)}
	\end{multline}
	The barotropic index (25) is singular for positive values of n(n $\neq$ 1), as was the case in the general relativistic cosmology but there is a significant difference here ; Type V singularities will not occur for all positive values of n besides $n=1$. For $ 0<n<0.75$, the w parameter remains non-singular which was not the case in the general relativistic scenario \cite{Dabrowski:2012eb}. Thus we find that for the ansatz (23), the f(R) gravity model described by (4) can have some significant differences in the conditions for the occurrence of type V singularities from the usual GR cosmology.   
	\section{Concluding remarks}
	In this paper, we have explored the conditions in which w-singularities can form in non-standard cosmological scenarios. While singularities of the types I-IV have been vividly discussed in the literature in many non-standard cosmological settings, type V or w-singularities have not yet been discussed in the same regards. Hence in this paper we studied the status quo of Type V singularities in a variety of different cosmological backgrounds. We considered the String theory inspired RS-II Braneworld cosmology, an F(R) gravity cosmology which gives viable late time acceleration and we also cosmologies due to modified area-entropy relations, generalized uncertainty principles, holographic renormalization and Chern-Simons gravity, all of which can be coincidentally described by the same modified Friedmann equation. We first considered a power series ansatz for the scale factor which was previously shown to give w-singularities and arrived at the rather surprising result that type V singularities will occur in exactly the same conditions in all these vividly different cosmological backgrounds as they did in the usual general relativistic cosmology in this scenario. The same constraint on the expansion for the scale factor which gave w-singularities in a general relativistic cosmology can give w-singularities in these non-standard cosmologies. We then considered an exponential ansatz for the scale factor which was also shown to give w-singularities and showed that while type V singularities occur in the same conditions in the RS-II Braneworld and Chern-Simons cosmologies as they did in the standard cosmology, the occurrence condition in f(R) gravity is a bit different. While in  type V singularities other cosmologies occur for all positive values of the parameter n besides $n=1$, in the f(R) gravity case they do not occur for $0<n<0.75$ besides the usual case of n=1. Thereby, the f(R) gravity case is the only one which shows a small but significant change in the conditions through which type V singularities can occur when one considers the exponential form for the scale factor. To conclude, we have shown that the conditions for the occurrence of w-singularities remain completely unscathed in a large class of non-standard cosmological backgrounds when one considers a power series expansion for the scale factor and the same can almost be said for the exponential scale factor, with the f(R) gravity case providing a noticeable exception here. One thing which we would like to note in passing is that all the cosmological scenarios that we undertook here were bore out of canonical actions and so one interesting future endeavour could be to see whether the conditions for the occurrence of w-singularities will change (if they do occur at all) if one considers non-canonical approaches to cosmology and in particular towards late-time evolution and dark energy.

	\section*{Acknowledgements}
	
	The work by MK has been supported by the grant of the Russian Science Foundation No-18-12-00213-P https://rscf.ru/project/18-12-00213/ and performed in Southern Federal University. The authors would also like to thank Sergei Odintsov, Shinichi Nojiri and Leonardo Fernandez-Jambrina for discussions and comments on cosmological singularities. We would also like to thank the anonymous reviwer for their insightful comments on the manuscript.

	\bibliography{JSPJMJ11.bib}
	
	\bibliographystyle{unsrt}

\end{document}